\documentclass[usenatbib]{mn2e}

\usepackage[dvips]{graphicx}
\usepackage{lscape}
\usepackage{txfonts}
\usepackage{url}
\usepackage{multirow}
\usepackage{times} 

\title{The merger history of massive spheroids since $z\sim$1 is size independent}

\author[L.A. D\'iaz-Garc\'ia et al.]{L.A. D\'iaz-Garc\'ia$^1$ \thanks{E-mail:diaz@cefca.es}, E. M\'armol-Queralt\'o$^{2,3}$, I. Trujillo$^{2,3}$, A.J. Cenarro$^{1}$, C. L\'opez-Sanjuan$^{1}$,\and P.G. P\'erez-Gonz\'alez$^{4,5}$, G. Barro$^{6}$ \\
$^{1}$Centro de Estudios de F\'{i}sica del Cosmos de Arag\'on (CEFCA), Plaza San Juan, 1, Planta-2, E-44001, Teruel, Spain\\
$^{2}$Instituto de Astrof\'{\i}sica de Canarias, c/ V\'{\i}a L\'actea s/n, 
E-38205, La Laguna, Tenerife, Spain\\
$^{3}$Departamento de Astrof\'{\i}sica, Universidad de La Laguna, E-38205, La 
Laguna, Tenerife, Spain\\
$^{4}$Departamento de Astrof\'{\i}sica, Facultad de CC. F\'{\i}sicas, Universidad Complutense de Madrid,
E-28040, Spain\\
$^{5}$Associated Astronomer at Steward Observatory, The University of Arizona\\
$^{6}$UCO/Lick Observatory, University of California, Santa Cruz, CA 95064}

\begin{document}

\date{}

\pagerange{\pageref{firstpage}--\pageref{lastpage}} \pubyear{2013}

\maketitle

\label{firstpage}

\begin{abstract}
Using a compilation of 379 massive (stellar mass M $\gtrsim$ 10$^{11}\,$M$_\odot$) spheroid-like galaxies from the near-infrared Palomar/DEEP-2 survey, we have probed, up to $z\sim$1, whether the presence of companions depends on the size of the host galaxies. We have explored the presence of companions with mass ratios down to 1:10 and 1:100, with respect to the central massive galaxy, and within a projected distance of 30, 50 and 100 kpc of these objects. We find evidence for these companions being equally distributed around both compact and extended massive spheroids. This finding suggests that, at least since $z\sim$1, the merger activity in these objects is rather homogeneous across the whole population and its merger history is not affected for the size of the host galaxy. Our result could indicate that both compact and extended massive spheroid-like galaxies are growing in size at the same rate.
\end{abstract}


\section{Introduction}\label{sec:introduction}

At a fixed stellar mass, the size of low redshift massive early-type galaxies (stellar mass M $\gtrsim$ 10$^{11}\,$M$_\odot$) is found to be a factor of two larger than their counterparts at $z\sim1$ \citep[e.g.,][]{Trujillo2004, Mcintosh2005, Trujillo2006, Trujillo2007,Buitrago2008,Newman2012}. In addition, the number density of such compact massive galaxies have decreased since that redshift \citep{Cassata2011}, with compact galaxies being extremely rare in the local Universe \citep{Trujillo2009,Taylor2010} and having young ages ($1-2$ Gyr, \citealt{Trujillo2009,Trujillo2012,Ferre-Mateu2012}), thus discarding them as the relics of the compact high-z galaxies. How have the compact massive galaxies evolved in size to occupy the present-day distribution? Having discarded major merging as the only mechanism to make galaxies grow in size since that epoch \citep[e.g.,][]{Lopez-Sanjuan2012}, two alternative ideas have been suggested: the puffing up model \citep{Fan2008, Fan2010, Damjanov2009} and the minor merging scenario \citep{Naab2009, Hopkins2009, Quilis2012}.

In the puffing-up model, galaxies grow in size by removing enormous quantities of gas by the effect of an AGN or by supernovae explosions at the early assembly of the spheroidal galaxies. Based on the analysis of the stellar population of the local and high-$z$ spheroidal galaxies, \cite{Trujillo2011} concluded that the evolution in size is independent of the stellar age. This has been one of the main argument against the puffing-up scenario. In the minor merging model, the  size evolution observed among the massive spheroid-like galaxy population is mainly caused by the continuous bombardment of smaller pieces  into the main objects. Recent studies have tried to quantify the impact of the observed merger rates in the size growth of massive galaxies \citep{Lopez-Sanjuan2012,Newman2012,Bluck2012}. These studies focus in the evolution of the average size--mass relation. To go further in this topic, not only the average, but also the intrinsic dispersion of the size--mass relation should be confronted against the observed merger histories. Observations find a nearly constant, even decreasing, dispersion since $z \sim 1.5$ \citep{Trujillo2007,Cassata2011,Newman2012}, while cosmological simulations suggest that merging tend to increase the dispersion of the size--mass relation with cosmic time \citep[e.g.,][]{Nipoti2012}. A fundamental ingredient in such model versus observation comparison is the variation of the merger rate with size at a given stellar mass, e.g., if compact galaxies have higher merger activity, they will evolve faster across the size--mass relation. In this paper we explore whether merging is acting in all the galaxies of a given stellar mass with the same probability or whether it is more frequent for galaxies with smaller sizes, which are quite rare in the local universe in comparison with the compact ones at higher redshift. Answering this question is key to understand how the local stellar mass-size relation has been built.

As a proxy to measure the minor merging activity in the massive spheroid galaxies since $z\sim1$ we study the frequency of companions around the massive galaxies since that epoch. We are working under the assumption that these companions will eventually be accreted over the main galaxy. In particular, we study whether the companions are preferentially located around galaxies with a specific size in the stellar size--mass relation or whether they are homogeneously distributed among the galaxy population independently of their stellar mass and size. We have focused our analysis up to $z\sim1$ which is approximately the redshift where our data, already presented in \citet[hereafter MQ12]{Marmol2012}, allow us to explore with completeness the presence of companions with stellar mass ratios between the central massive galaxy ($M_{\rm central}$) and its companion ($M_{\rm com}$) down to $0.01<M_{\rm com}/M_{\rm central}<1$ (1:100) around our sample of massive (stellar mass $M \gtrsim 10^{11}$M$_\odot$) galaxies. 

This paper is organized as follows. In Section~\ref{sec:data} we present the two samples used in this work: the sample of central massive spheroid-like galaxies with a brief description of the Palomar/DEEP-2 survey and the sample of companions, from The Rainbow Database, around them. We explain the companion detection process in Section~\ref{sec:selection}. We present the analysis of the data in Section~\ref{sec:analysis} together with the contamination corrections due to the uncertainties in the photometric redshifts. Finally, in Section~\ref{sec:conclusion}, we present the conclusions of our findings and a brief summary of this paper. In this paper we adopt a standard $\Lambda$CDM cosmology, with $\Omega_{\rm m} = 0.3$, $\Omega_{\rm \Lambda} = 0.7$, $H_0 = 100h$ km~s$^{-1}$~Mpc$^{-1}$ and $h=0.7$.


\section{The data}\label{sec:data}

To study whether the size of the massive spheroids is relevant for their merger history, we use a complete, mass selected, large catalogue of massive galaxies. We explore which one of these objects have companions which would be able to merge with their central massive galaxy. Thus, our analysis is based on two different datasets: a catalogue with central massive spheroids and another sample containing their companion galaxies.

As the reference catalogue for the central massive spheroid-like galaxies we have used the compilation of \citet[hereafter T07]{Trujillo2007}. The {\it K$_s$}-band imaging from the Palomar Observatory Wide-field Infrared (POWIR)/DEEP-2 survey \citep{Davis2003, Bundy2006, Conselice2007} was used to define a sample of 831 massive galaxies (stellar mass $M>10^{11}$M$_\odot$) up to $z=2$ located over $\sim$710 arcmin$^2$ in the Extended Groth Strip (EGS). Furthermore, these objects were imaged with the Advanced Camera for Surveys (ACS) from Hubble Space telescope (HST) in the {\it F606W} and {\it F814W} bands, with the CFH12K camera from the Canada--France--Hawaii Telescope in the \emph{B}, \emph{R} and \emph{I} bands, and in \emph{J} and {\it K$_s$} bands from the Palomar 5-m telescope. T07 present massive galaxies with spectroscopic redshifts that have been supplemented with photometric redshifts with an accuracy of $\Delta z/(1+z)\sim 0.07$. Stellar masses were estimated using a \citet{Chabrier2003} initial mass function. T07 estimated circularized half-light radius ($r_{\rm e}$) and S\'ersic indices $n$ \citep[][]{Sersic1968} for all the galaxies in the central sample. The criteria used to identify the massive spheroid-like objects is based on their S\'ersic indices. The S\'ersic index can be used to make a reliable morphological classification of galaxies since it measures the shape of surface brightness profile \citep{Andredakis1995}. To obtain a reliable sample of bulge-dominated galaxies and to exclude the late-type galaxies with a bright nucleus, we select the galaxies which have S\'ersic indices larger than 2.5 (see fig.~1 in \citealt{Ravindranath2004}).

To compile the sample of companion galaxies around our massive spheroid-like objects we have used the Extended Groth Strip IRAC-selected galaxy sample from the Rainbow Cosmological Database\footnote{\url{https://rainbowx.fis.ucm.es/Rainbow_Database/}} published by \citet[][see also \citealt{Perez-Gonzalez2008}]{Barro2011a}. This database provides spectral energy distributions (SEDs) ranging from the UV to the MIR regime plus well-calibrated and reliable photometric redshifts and stellar masses \citep{Barro2011b}. In general, the sample of companions has photometric redshifts from the Rainbow Database. We refer to this resulting sample as the Rainbow catalogue.

\begin{figure*}
\centering
\includegraphics[width=1.\textwidth,clip=True]{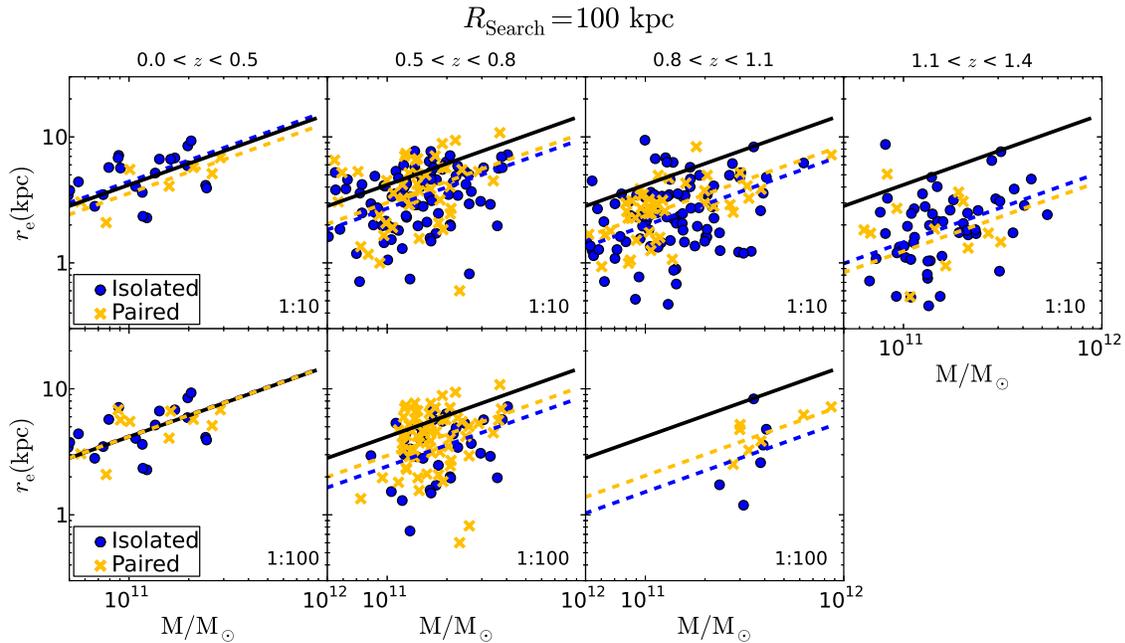}
\caption{Stellar size--mass relation of spheroid-like galaxies in different redshift bins. The different panels show the distribution of paired (yellow crosses) and isolated (blue dots) galaxies down to mass ratios of 1:10 (upper panels) and 1:100 (lower panels), in all cases within a projected distance of $100$ kpc. The black line represents the local stellar size--mass relation \citep{Shen2003} for spheroid-like galaxies. For every redshift bin, dashed yellow and blue lines show the best-fitting to the distribution of paired and isolated galaxies respectively (see Section~\ref{sec:analysis} for more details).}
\label{fig:sample100}
\end{figure*}

\begin{figure*}
\centering
\includegraphics[width=1.\textwidth,clip=True]{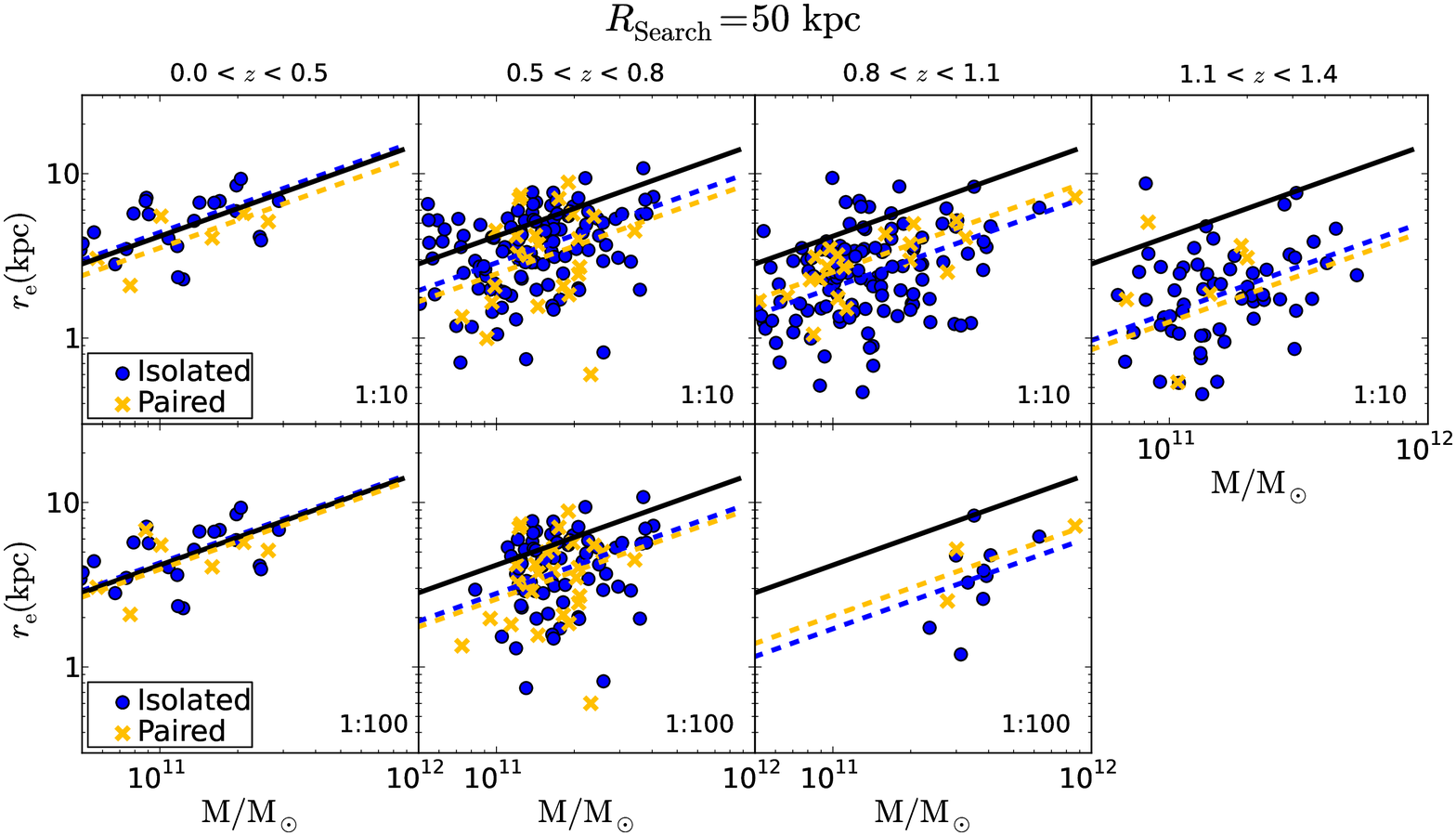}
\caption{Same as Fig.~\ref{fig:sample100} but using a search radius of $50$ kpc for the companion detection process (see Section~\ref{sec:selection}).}
\label{fig:sample50}
\end{figure*}

\begin{figure*}
\centering
\includegraphics[width=1.\textwidth,clip=True]{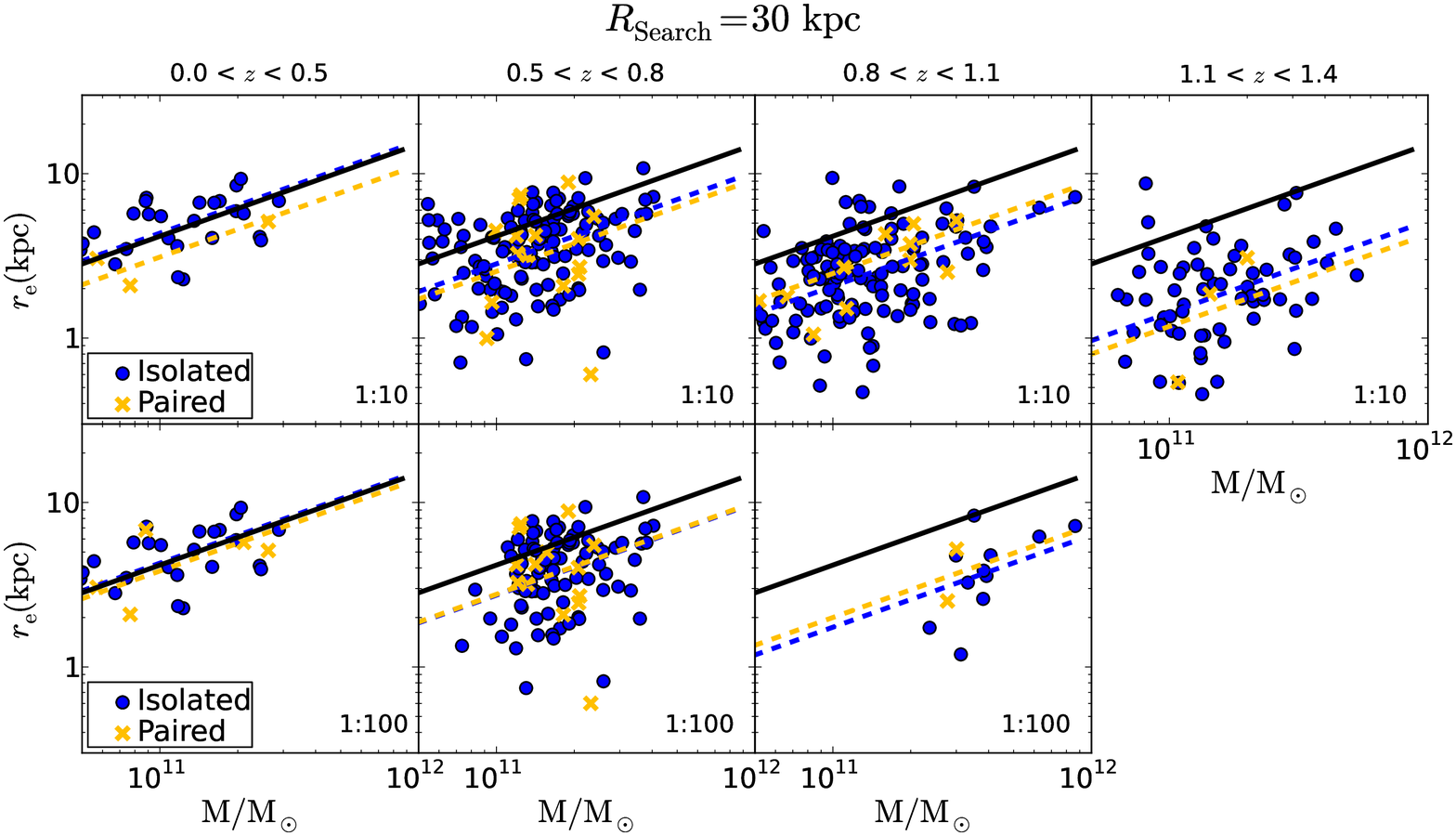}
\caption{Same as Fig.~\ref{fig:sample100} but using a search radius of $30$ kpc for the companion detection process (see Section~\ref{sec:selection}).}
\label{fig:sample30}
\end{figure*}

In order to build a sample of central massive galaxies with the best estimations of redshifts and stellar masses we have followed the same criteria as in MQ12: if a given central galaxy has a spectroscopic redshift in the Rainbow catalogue, both the redshift and the stellar mass are taken from this catalogue. Otherwise, spectroscopic redshifts and stellar masses are taken from T07. If the central galaxy has no spectroscopic redshift in any of the two above data sets, we take the Rainbow photometric redshift as long as it is in agreement with the photometric redshift of T07. More precisely, we impose that the difference between the photometric redshifts in both catalogues has to be smaller than $0.070$ at $0.0<z<0.5$, $0.061$ at $0.5<z<1.0$, and $0.083$ at $z>1.0$. When a larger difference is found, the central galaxy is rejected for this study.

To assure that the fraction of galaxies with companions along our explored redshift range is not biased by the stellar mass completeness limit of the Rainbow catalogue, we only keep in the central sample those galaxies that are at least 10 (100) times more massive than the stellar mass limit of the companion sample at each redshift. The stellar mass limit (75 per cent complete) of the Rainbow catalogue at each redshift is provided in \citet{Perez-Gonzalez2008} and it ranges from $M \gtrsim 10^{8.5}$M$_\odot$ at $z\sim0.2$ to $M \gtrsim 10^{10}$M$_\odot$ at $z\sim1.2$ (see fig.~4 in \citealt{Perez-Gonzalez2008}). These stellar mass limits correspond to the stellar mass of a passively evolving galaxy formed in a single instantaneous burst of star formation occurred at $z\sim \infty$ and having a $\rm 3.6\mu m$ flux equal to the 75 per cent completeness level in the IRAC galaxy sample ([3.6]$\sim24.75$m$_{\rm AB}$). In the redshift range $0<z<1.4$ we can probe companions with a mass fraction compared to their central objects (stellar mass $M$ $\gtrsim$ 10$^{11}$M$_\odot$) down to 1:10. There are finally 379 massive spheroid-like galaxies that meet all the above criteria, out of which 239 have spectroscopic redshifts and 140 have photometric redshifts. In the redshift range  $0<z<1.1$ we are able to explore the presence of companions down to a mass ratio of 1:100. The number of massive spheroid-like galaxies for which this study can be conducted is 145 (107 with spectroscopic redshifts and 38 with photometric redshifts).


\section{Detection of companions}\label{sec:selection}
The process and criteria to search for companions around massive spheroid-like galaxies is based in MQ12. A galaxy from the Rainbow catalogue is considered as a potential companion, if the redshift difference between the object and the central host is less than $1\sigma$ uncertainty, within a projected radial distance $R_{\rm search}$ around the central object. Usual values of $R_{\rm search}$ in the literature ranges from 30 kpc to 150 kpc in our reference cosmology {\citep[$h=0.7$, e.g.][]{Patton2000,Bell2006,Marmol2012,Bluck2012}}. With these $R_{\rm search}$ values, the selected pairs will merge in a relatively short time-scale \citep[t $\lesssim 2.5$ Gyr,][]{Lotz2010}. In the following we explore three different search radii, $R_{\rm search} = 30$, $50$ and $100$ kpc. Larger search radius increases the background contamination and the results become very uncertain. Finally, we only consider companion objects within a mass range of $0.1<M_{\rm com}/M_{\rm central}<1$ if we explore our sample up to $z=1.4$ and companions within $0.01<M_{\rm com}/M_{\rm central}<1$ if we constrain our analysis to massive galaxies with $z<1.1$. Hereafter, we refer to the massive spheroid-like galaxies with companions as paired galaxies and the massive spheroid-like galaxies without companions as isolated galaxies. These concepts have been defined for this work, so the definitions may differ with other publications.

It is clear that the main source of contamination in this sort of studies comes from the uncertainties in the photometric redshifts. Due to the background/foreground contamination, there exists a fraction of fake paired galaxies with this method of companion detection (see Section~\ref{sec:correction1}). In addition, there also exists a fraction of fake isolated galaxies given that the redshift difference between the host galaxy and the potential candidate is set to be within a $1\sigma$ uncertainty (see Section~\ref{sec:correction2}). Note that these two effects can be constrained and corrected only statistically. Since the study in the present paper is performed over individual galaxies and it is not possible to state whether the companion of a particular massive central galaxy is real or a contaminant, we devote Section~\ref{sec:reliability} to probe in detail the impact of these two effects in our results as well as to account statistically for them by means of Monte Carlo simulations.


\section{Analysis of the data}\label{sec:analysis}

The main goal of the present work is to address whether the presence of companions, with mass ratios down to 1:10 and 1:100, depends on the size of massive spheroid-like galaxies. We split our sample in different redshift bins to analyse the cosmic time evolution of this dependence. In Figs.~\ref{fig:sample100},~\ref{fig:sample50}, and~\ref{fig:sample30} we show the stellar size--mass relation of our sample of central massive spheroidal-like galaxies at different redshift bins and search radii $R_{\rm search} = 30, 50,$ and $100$ kpc; according to the criteria described in Section~\ref{sec:selection}. A quick look to these plots suggests that the paired galaxies are distributed homogeneously through the stellar size--mass relation at all redshifts, independently of whether we explore the existence of companions in the 1:10 or in the 1:100 mass ratio range. We will now quantify this in more detail.

\cite{Shen2003} parametrized the size--mass relation of spheroid-like galaxies in the local Universe from the Sloan Digital Sky Survey \citep[SDSS,][]{Stoughton2002} with a power--law function, 
\begin{equation}\label{eq:size_mass}
r_e\,{\rm(kpc)} = b \left( \frac{M}{\rm M_{\odot}} \right)^a,
\end{equation}

finding $a=0.56$ and $b=2.88 \times 10^{-6}$\,kpc. Interestingly, some recent studies \citep[e.g.,][]{Damjanov2011,Mclure2013} have found that the size--mass relation for spheroid-like galaxies with stellar masses M $\gtrsim$ $10^{10.5}\,$M$_\odot$ does not change its slope, at least up to $z \sim 1.5$, and this slope is compatible with the obtained for \cite{Shen2003} in the local universe. For this reason, in what follows we will assume that $a$ is independent on redshift. To quantify whether there exists differences between the size distributions of paired and isolated galaxies in our sample, we fit the distributions in Figs.~\ref{fig:sample100},~\ref{fig:sample50}, and ~\ref{fig:sample30} using Eq.~(\ref{eq:size_mass}). Error-weighted least-squares linear fittings have been performed to the two galaxy samples in each redshift bin. We keep the power-law index $a$ fixed to the same value obtained by \cite{Shen2003} throughout the fitting process, while $b$ is left as a free parameter. To determine the best-fitting, we assumed that the accuracy of the half-light radius is $\sigma_{r_{\rm e}}/r_{\rm e} \sim 0.2$ and constant among our galaxy sample \citep[see][]{Trujillo2007}. In Figs.~\ref{fig:sample100},~\ref{fig:sample50}, and ~\ref{fig:sample30} the yellow and blue dashed lines show the results of the fittings for the distributions of paired and isolated galaxies, respectively, whereas the black line is the local size-mass relation obtained by \cite{Shen2003} in the local universe for early-type galaxies. In Table \ref{tab:slopes} we provided the best-fitting $b$ values together with their errors (confidence level of 68.3 per cent) for the distribution of paired ($b_{\rm pair}$) and isolated ($b_{\rm iso}$) galaxies. The ratio between the $b$ parameters of both populations is consistent with unity within the errors for every search radius and mass ratio, which implies that there exists no significant shift between the two distributions.

\begin{table*}
\centering
\caption{The best-fitting parameter $b$ (see Eq.~\ref{eq:size_mass}) of the stellar size--mass distribution for paired and isolated galaxies as a function of redshift, search radius and mass ratio.}
\begin{tabular}{lccccccc}
\hline\hline 
\multicolumn{8}{|c|}{Parameter $b$ $[10^{-6} \rm{kpc}]$} \\
\hline
\multirow{2}{*}{\bf Mass ratio $\bf{1:10}$}      & \multicolumn{2}{c}{\multirow{2}{*}{$R_{\rm search} = 30 \ {\rm kpc}$}}& \multicolumn{2}{c}{\multirow{2}{*}{$R_{\rm search} = 50 \ {\rm kpc}$}}& \multicolumn{2}{c}{\multirow{2}{*}{$R_{\rm search} = 100 \ {\rm kpc}$}} & \\
& & & & & & & \\
\hline
Redshift & $b_{\rm pair}$ & $b_{\rm iso}$ & $b_{\rm pair}$ & $b_{\rm iso}$ & $b_{\rm pair}$ & $b_{\rm iso}$ & Global\\
\hline
$0.0<z<0.5$ & $2.15 \pm 0.58 $ & $3.01 \pm 0.28 $ & $2.45 \pm 0.46 $ & $3.04 \pm 0.29 $ & $2.47 \pm 0.43 $ & $3.06 \pm 0.30 $ &   $ 2.90 \pm 0.25 $  \\
$0.5<z<0.8$ & $1.76 \pm 0.20 $ & $1.96 \pm 0.08 $ & $1.70 \pm 0.16 $ & $1.98 \pm 0.08 $ & $2.08 \pm 0.14 $ & $1.87 \pm 0.09 $ &   $ 1.93 \pm 0.07 $  \\
$0.8<z<1.1$ & $1.71 \pm 0.23 $ & $1.43 \pm 0.06 $ & $1.74 \pm 0.17 $ & $1.40 \pm 0.06 $ & $1.64 \pm 0.11 $ & $1.37 \pm 0.06 $ &   $ 1.45 \pm 0.05 $  \\
$1.1<z<1.4$ & $0.82 \pm 0.22 $ & $0.99 \pm 0.06 $ & $0.87 \pm 0.15 $ & $0.99 \pm 0.06 $ & $0.86 \pm 0.11 $ & $1.01 \pm 0.06 $ &   $ 0.98 \pm 0.05 $  \\
\hline
\multirow{2}{*}{\bf Mass ratio $\bf{1:100}$}      & \multicolumn{2}{c}{\multirow{2}{*}{$R_{\rm search} = 30 \ {\rm kpc}$}}& \multicolumn{2}{c}{\multirow{2}{*}{$R_{\rm search} = 50 \ {\rm kpc}$}}& \multicolumn{2}{c}{\multirow{2}{*}{$R_{\rm search} = 100 \ {\rm kpc}$}} & \\
& & & & & & & \\
\hline
Redshift & $b_{\rm pair}$ & $b_{\rm iso}$ & $b_{\rm pair}$ & $b_{\rm iso}$ & $b_{\rm pair}$ & $b_{\rm iso}$ & Global\\
\hline
$0.0<z<0.5$  & $2.65 \pm 0.55 $ & $2.96 \pm 0.28 $ & $2.71 \pm 0.47 $ & $2.97 \pm 0.29 $ & $2.89 \pm 0.42 $ & $2.91 \pm 0.31 $ & $ 2.90 \pm 0.25$\\
$0.5<z<0.8$  & $1.91 \pm 0.24 $ & $1.89 \pm 0.09 $ & $1.79 \pm 0.15 $ & $1.94 \pm 0.10 $ & $2.04 \pm 0.12 $ & $1.68 \pm 0.12 $ & $ 1.90 \pm 0.09$\\
$0.8<z<1.1$  & $1.39 \pm 0.46 $ & $1.21 \pm 0.17 $ & $1.42 \pm 0.38 $ & $1.18 \pm 0.17 $ & $1.41 \pm 0.25 $ & $1.05 \pm 0.20 $ & $ 1.23 \pm 0.16$\\
\hline
\end{tabular}
\label{tab:slopes}
\end{table*}

\begin{figure}
\centering
\includegraphics[width=1.\columnwidth,clip=True]{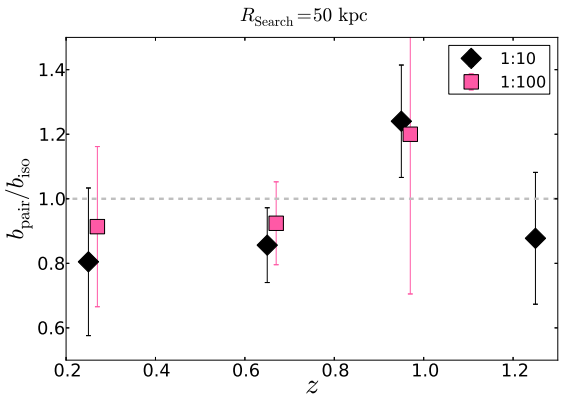}
\caption{The ratio of the $b$ parameters of the size--mass relations (Eq.~\ref{eq:size_mass}) for paired and isolated massive central galaxies in different redshift bins and mass ratios (1:10, black diamonds; 1:100, pink squares), The figure shows the case when the companions are explored using a search radius of 50 kpc. A ratio close to 1 implies that the average size of paired and isolated massive galaxies is the same.}
\label{fig:bsat_bnosat50}
\end{figure}

To confirm the above result we performed Kolmogorov-Smirnov and t-Student tests to the distributions of paired and isolated galaxies in the size--mass plane. To do that, we performed the fitting of Eq.~(\ref{eq:size_mass}) to the whole central galaxy sample (see global parameters in Table~\ref{tab:slopes}), for every redshift and mass range, and studied the distances to this relation for the two galaxy cases. In Table~\ref{tab:tests} we show the Kolmogorov-Smirnov and t-Student estimators obtained for our sample of central massive galaxies, denoted $E_{\rm KS}$ and $E_{\rm TS}$ respectively, as well as their limiting values for confidence levels of 95 per cent of the two distributions, taking into account the degrees of freedom of each subsample. In all cases the estimators are smaller than the limiting values, consequently we infer that the two distributions are statistically indistinguishable at the 95 per cent confidence level. This implies that the size--mass distributions of paired and isolated galaxies are not statistically different.

\begin{figure*}
\centering
\includegraphics[width=1.\textwidth,clip=True]{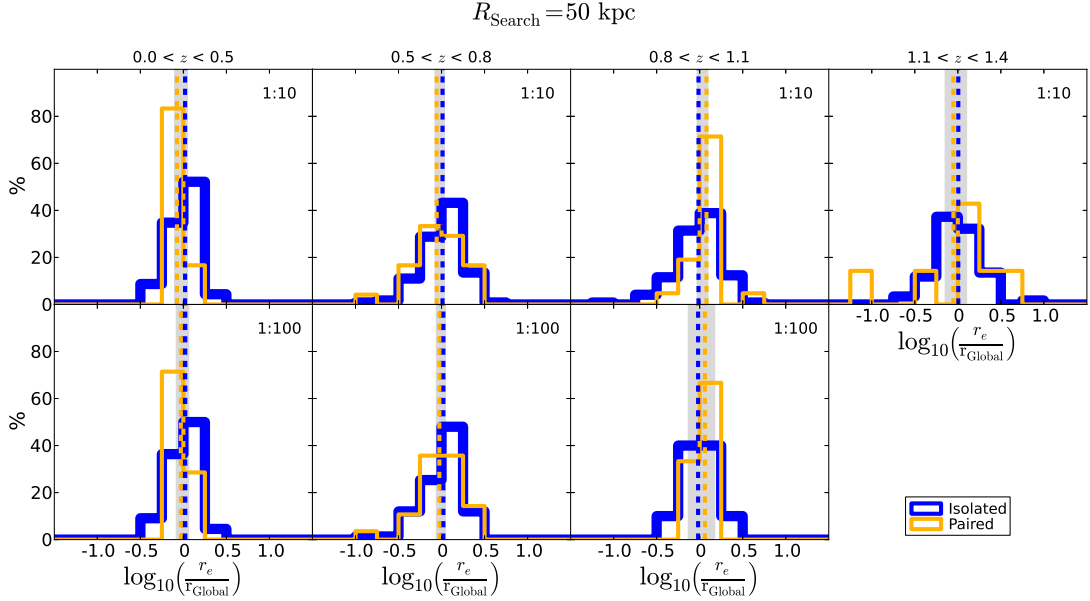}
\caption{Histograms of the distances of paired (yellow) and isolated (blue) galaxies to the size--mass relation (Eq.~\ref{eq:size_mass}) for the whole central galaxy sample, using a search radius of 50 kpc and mass ratios of 1:10 (upper panels) and 1:100 (lower panels). Dashed yellow and blue lines show the mean values of the paired and isolated distributions, respectively. The shadow area is the difference between the means of two distributions for a confidence level of 95 per cent with the same degrees of freedom for a t-Student test. In all cases, the means of the two populations are statistically the same.}
\label{fig:hist50}
\end{figure*}

\begin{figure*}
\centering
\includegraphics[width=1.\textwidth,clip=True]{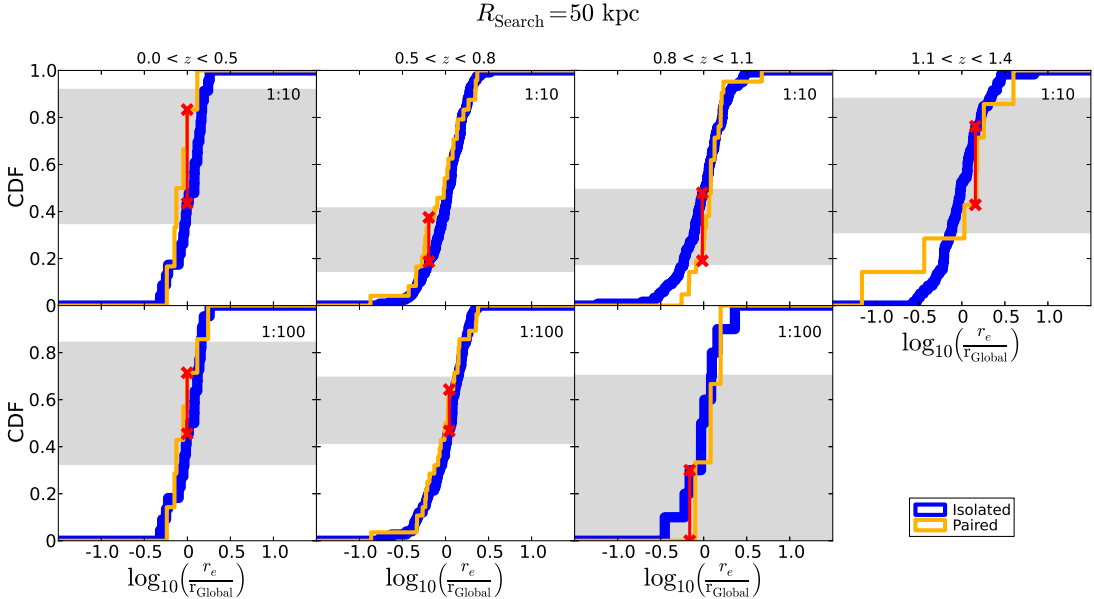}
\caption{CDFs of the distances of paired (yellow) and isolated (blue) galaxies to the size--mass relation (Eq.~\ref{eq:size_mass}) for the whole central galaxy sample, using a search radius of 50 kpc and mass ratios of 1:10 (upper panels) and 1:100 (lower panels). Red line illustrates the largest distance between the CDFs of paired and isolated galaxies. The shadow area  is the difference between the CDFs of two distributions for a confidence level of 95 per cent with the same degrees of freedom for a Kolmogorov-Smirnov test. In all cases, the shifts between the CDFs are statiscally compatible.}
\label{fig:cdf50}
\end{figure*}

The most representative search radius for this study is 50 kpc as both the background/foreground contamination and the detection contamination are fairly low ($\sim 10$ per cent in both cases) and the number of paired galaxies ensures good statistics. For a search radius of 50 kpc, Fig.~\ref{fig:bsat_bnosat50} illustrates that the ratio between the $b$ parameters of both populations (paired and isolated galaxies) is consistent with unity within the errors, hence implying that there is no significant shift between the two distributions. Figure~\ref{fig:hist50} illustrates the histograms of the distances of paired (yellow) and isolated (blue) galaxies to the fitting of the size--mass relation (Eq.~\ref{eq:size_mass}) of the whole central galaxy sample for a search radius of $50$ kpc. Dashed yellow and blue lines show the mean values of the paired and isolated distributions respectively. For a t-Student test, the shadow area is the difference between the means of two distributions for a confidence level of 95 per cent with the same degrees of freedom. In all cases, the means of the two populations are statistically the same for a confidence level of 95 per cent in the t-Student test. The Fig.~\ref{fig:cdf50} shows the Cumulative Distribution Function (CDF) of the histograms in Fig.~\ref{fig:hist50}. The vertical red line is the largest difference between the CDFs of the paired and isolated galaxies. For a Kolgomorov-Smirnov test, the shadow area is the difference between the CDFs of two distributions for a confidence level of 95 per cent with the same degrees of freedom. In every panel, the red line is smaller than the shadow area so that both distributions (paired and isolated) are statistically the same in the Kolmogorov-Smirnov test. We have obtained similar results for the search radii $R_{\rm search} = 30$ and $100$ kpc.

\subsection{Systematic effect analysis}

These results have been obtained by comparing the distributions of paired and isolated galaxies, but neither the background/foreground contamination nor the fake isolated galaxy contamination were taking into account since we can not correct for these effects individually. Both effects are studied in Section~\ref{sec:correction1} and Section~\ref{sec:correction2} respectively. Finally, Section~\ref{sec:reliability} is devoted to check the impact of these contamination sources in our results.

\subsubsection{Background/foreground contamination correction}\label{sec:correction1}

With the method of companion detection described in Section~\ref{sec:selection}, MQ12 shows that there exists an excess in the number of paired galaxies due to the photometric redshift uncertainties (background/foreground contamination). Since we could be taking companion candidates when they are not really linked gravitationally to the host galaxy, leading to fake paired galaxies, such redshift uncertainties are the main source of contamination. Consequently, it is necessary to estimate statistically the background/foreground contamination and check to what extent our results may be affected.

We have followed the same statistical process in MQ12 to estimate the background/foreground contamination. This method consists on placing mock massive galaxies randomly in the volume of the Rainbow catalogue. The number of mock galaxies that are in every redshift bin is the same than in our observed sample and the mock galaxy parameters are the same than the massive spheroid-like galaxy parameters, such as redshifts and stellar masses. Once the mock galaxies are situated in the Rainbow catalogue volume, we apply the companion detection process explained above (including the $1\sigma$ uncertainty in the estimation of the redshifts) and compute the fraction of these mock galaxies having companions, $F_{\rm pair,mock}$, around them for search radii $R_{\rm search}=30,50$ and $100$ kpc, in the mass ratio 1:10 and 1:100. To get a robust count of the paired mock galaxies, we repeat this process one hundred thousand times. We reasonably assume that the average fraction of paired mock galaxies, $\overline{F}_{\rm pair,mock}$, corresponds to the fraction of galaxies affected by the background/foreground contamination.

The observed fraction of paired galaxies, $F_{\rm pair,obs}$, is the sum of the real fraction of paired galaxies, $F_{\rm pair}$, plus the fraction of isolated galaxies affected by the background/foreground contamination, $(1-F_{\rm pair})\times \overline{F}_{\rm pair,mock}$. Thus, the true fraction of paired galaxies after the statistical correction is

\begin{equation}
F_{\rm pair} = \frac{F_{\rm pair,obs}-\overline{F}_{\rm pair,mock}}{1-\overline{F}_{\rm pair,mock}}.
\label{eq:fraction}
\end{equation}

The fraction of fake paired galaxies as a consequence of our companion detection process is

\begin{equation}
C_{\rm pair}= \frac{F_{\rm pair,obs}-F_{\rm pair}}{F_{\rm pair,obs}}.
\label{eq:contamination1}
\end{equation}

The simulation results, shown in Table~\ref{tab:contamination}, indicate that up to $\sim$4, 9, 24 per cent ($\sim$6, 14, 26 per cent) of the observed paired galaxies in the mass ratio 1:10 (1:100) actually have no companions within a search radius of 30, 50 and 100 kpc.

\subsubsection{Fake isolated galaxies due to the $1\sigma$ uncertainty condition}\label{sec:correction2}

In the companion detection process, the candidates have been constrained to have a difference between redshifts lower than 1$\sigma$ uncertainty with the central massive galaxy. Assuming that both, the central galaxy and its companion, have photometric redshifts, we expect to miss $F_{\sigma} \sim 30$ per cent of the companions in our search. MQ12 estimated that allowing for a difference of up to 2$\sigma$, instead of 1$\sigma$ in the search of companions, the background/foreground contamination effects rise $\sim50$ per cent whereas the fraction of paired galaxies changes less than $30$ per cent, as expected. Thus, the 1$\sigma$ condition is the optimal one for merger fraction studies, as shown by MQ12. 

Since in the present paper we study individual systems, we have taken into account the fraction of fake isolated galaxies due to the 1$\sigma$ condition. We may expect that the total fraction of massive spheroid-like galaxies with a companion, $F_{\rm pair,T}$, is 
\begin{equation}
F_{\rm pair,T} = \frac{F_{\rm pair}}{1-F_{\sigma}}.
\label{eq:total}
\end{equation}

Being $N_{\rm central}$ the number of massive galaxies, $(F_{\rm pair,T} - F_{\rm pair}) \times N_{\rm central}$ is the number of fake isolated galaxies due to the 1$\sigma$ uncertainty. The fraction $C_{\rm iso}$ of observed isolated galaxies that truly have a companion is 

\begin{equation}
C_{\rm iso} = \frac{F_{\rm pair,T}-F_{\rm pair}}{1-F_{\rm pair,obs}}.
\label{eq:contamination2}
\end{equation}

Because $F_{\sigma} \sim 30$ per cent, we estimate that $C_{\rm iso}$ (see Table~\ref{tab:contamination}) is lower than 4, 7, 14 per cent (7, 14, 46 percent) for search radius of 30, 50 and 100 kpc in the mass ratio 1:10 (1:100).

\subsubsection{Result reliability}\label{sec:reliability}

Due to the photometric redshift uncertainties, we must check whether the contamination in the companion detection can be affecting our results. Since we cannot correct for the contamination of central galaxies individually, we adopt a statistical Monte Carlo approach, as a sanity check, to ensure that the background/foreground contamination and the $1\sigma$ assumption are not compromising our results. First, we randomly move central paired galaxies from the observed sample to the isolated sample, to reach the expected number of fake paired galaxies after the contamination correction according to Table~\ref{tab:contamination}. Secondly, we randomly move observed isolated galaxies to the set of paired galaxies, independently of their sizes or masses, until the expected numbers of fake isolated galaxies presented in Table~\ref{tab:contamination} are recovered. Since the errors in the detection of companions are fully linked to the errors in the determination of the redshifts, we can reasonably assume that the fake paired and fake isolated galaxies will be randomly located around the size--mass relation, independently of the mass or size. Then, we apply Kolmogorov-Smirnov and t-Student tests over the new samples of paired and isolated galaxies. We repeat this process one million times and determine the fraction of cases for which both distributions are undistinguishable, according to the above statistical tests.

\begin{table*}
\centering
\caption{Test estimators obtained of the Kolmogorov-Smirnov and t-Student tests, denoted $E_{\rm KS}$ and $E_{\rm TS}$ respectively. We present the values of the estimators for a confidence level of 95 per cent, denominated $E_{\rm KS}(95\%)$ and $E_{\rm TS}(95\%)$, taking into account the degree of freedoms of each subsample.}
\begin{tabular}{lcccccccc}
\hline\hline
\multicolumn{9}{c}{Test estimators}\\
\hline
& \multicolumn{4}{c}{\multirow{2}{*}{\bf Mass ratio $\bf{ 1:10}$}} & \multicolumn{4}{c}{\multirow{2}{*}{\bf Mass ratio $\bf{1:100}$}}  \\
&&&&&&&&\\
\hline
& \multicolumn{2}{c}{Kolmogorov-Smirnov}  & \multicolumn{2}{c}{t-Student} & \multicolumn{2}{c}{Kolmogorov-Smirnov}  & \multicolumn{2}{c}{t-Student}  \\
Redshift & $E_{\rm KS}$ & $E_{\rm KS}(95\%)$ & $E_{\rm TS}$ & $E_{\rm TS}(95\%)$ & $E_{\rm KS}$ & $E_{\rm KS}(95\%)$ & $E_{\rm TS}$ & $E_{\rm TS}(95\%)$ \\
\hline
\multirow{2}{*}{R$_{\rm search} = 30 \ {\rm kpc}$} & & & & & & \\
&&&&&&&& \\
\hline
$0.0<z<0.5$ & $0.538$ & $0.720$ & $1.502$ & $2.052$ & $0.342$ & $0.602$ & $0.588$ & $2.052$ \\
$0.5<z<0.8$ & $0.176$ & $0.344$ & $0.715$ & $1.977$ & $0.203$ & $0.372$ & $0.069$ & $1.984$ \\
$0.8<z<1.1$ & $0.204$ & $0.386$ & $0.957$ & $1.977$ & $0.409$ & $0.864$ & $0.359$ & $2.201$ \\
$1.1<z<1.4$ & $0.270$ & $0.712$ & $0.444$ & $1.998$ &         &         &         &         \\
\hline
\multirow{2}{*}{R$_{\rm search} = 50 \ {\rm kpc}$} & & & & & & \\
&&&&&&&& \\
\hline
$0.0<z<0.5$ & $0.408$ & $0.569$ & $1.284$ & $2.052$ & $0.298$ & $0.538$ & $0.550$ & $2.052$ \\
$0.5<z<0.8$ & $0.189$ & $0.293$ & $1.261$ & $1.977$ & $0.176$ & $0.291$ & $0.649$ & $1.984$ \\
$0.8<z<1.1$ & $0.289$ & $0.308$ & $1.467$ & $1.977$ & $0.300$ & $0.749$ & $0.563$ & $2.201$ \\
$1.1<z<1.4$ & $0.334$ & $0.505$ & $0.458$ & $1.998$ &         &         &         &         \\
\hline
\multirow{2}{*}{R$_{\rm search} = 100 \ {\rm kpc}$} & & & & & & \\
&&&&&&&& \\
\hline
$0.0<z<0.5$ & $0.448$ & $0.538$ & $1.346$ & $2.052$ & $0.184$ & $0.486$ & $0.045$ & $2.052$ \\
$0.5<z<0.8$ & $0.172$ & $0.235$ & $1.036$ & $1.977$ & $0.228$ & $0.262$ & $1.773$ & $1.984$ \\
$0.8<z<1.1$ & $0.184$ & $0.235$ & $1.632$ & $1.977$ & $0.524$ & $0.675$ & $1.133$ & $2.201$ \\
$1.1<z<1.4$ & $0.235$ & $0.393$ & $0.742$ & $1.998$ &         &         &         &         \\
\hline
\end{tabular}
\label{tab:tests}
\end{table*}

\begin{table*}
\centering
\caption{Bias in the determination of paired and isolated galaxies due to the photometric redshift uncertainties. For each redshift range we present the number of massive galaxies $N_{\rm central}$, the observed fraction of paired massive galaxies  $F_{\rm pair,obs}$ and the fraction of paired galaxies after the contamination correction $F_{\rm pair}$ with their errors. We show the overestimation in the number of paired galaxies due to the redshift uncertainties with this method of companion detection $C_{\rm pair}$. Due to the assumption of a 1$\sigma$ uncertainty criteria, we also present the fraction of fake isolated galaxies, $C_{\rm iso}$.}
\begin{tabular}{lcccccccccccc}
\hline\hline
& \multicolumn{5}{c}{\bf Mass ratio ${\bf 1:10}$}& \multicolumn{5}{c}{\bf Mass ratio ${\bf 1:100}$} \\
\hline
Redshift range &$N_{\rm central}$ & $F_{\rm pair,obs}$ & $F_{\rm pair}$ & $C_{\rm pair}$ & $C_{\rm iso}$ &$N_{\rm central}$ & $F_{\rm pair,obs}$ & $F_{\rm pair}$ & $C_{\rm pair}$ & $C_{\rm iso}$\\
\hline
\multirow{2}{*}{$R_{\bf \rm search} = 30 \ {\rm kpc}$} & & & & & & & & & & &\\
& & & & & & & & & &\\
\hline
$0.0 < z < 0.5$	&	$29$	&	$0.100$	&	$0.097 \pm 0.010$ & $0.03$	& $0.05$	&$29$	&	$0.200$	&	$0.192 \pm 0.019$ &	$0.04$ & $0.10$ \\
$0.5 < z < 0.8$	&	$142$	&	$0.113$	&	$0.110 \pm 0.005$ & $0.03$	& $0.05$	&$103$	&	$0.136$	&	$0.127 \pm 0.010$ &	$0.07$ & $0.06$ \\
$0.8 < z < 1.1$	&	$142$	&	$0.085$	&	$0.082 \pm 0.004$ & $0.04$	& $0.04$	&$13$	&	$0.154$	&	$0.147 \pm 0.026$ &	$0.05$ & $0.07$ \\
$1.1 < z < 1.4$	&	$66$	&	$0.045$	&	$0.042 \pm 0.006$ & $0.07$	& $0.02$   &       &           &                     &    	    &         \\
\hline
\multirow{2}{*}{$R_{\rm search} = 50 \ {\rm kpc}$} & & & & & & & & &\\
& & & & & & & & & &\\
\hline
$0.2 < z < 0.5$	&	$29$	&	$0.20$	&	$0.19 \pm 0.02$  & $0.06$  & $0.10$	&$29$	&	$0.27$	&	$0.23 \pm 0.04$ & $0.13$	& $0.14$ \\
$0.5 < z < 0.8$	&	$142$	&	$0.17$	&	$0.16 \pm 0.01$  & $0.09$  & $0.08$	&$103$	&	$0.27$	&	$0.23 \pm 0.02$ & $0.14$	& $0.14$ \\
$0.8 < z < 1.1$	&	$142$	&	$0.15$	&	$0.14 \pm 0.01$  & $0.08$  & $0.07$	&$13$	&	$0.23$	&	$0.20 \pm 0.06$ & $0.14$	& $0.11$ \\
$1.1 < z < 1.4$	&	$66$	&	$0.11$	&	$0.09 \pm 0.01$  & $0.12$  & $0.04$ &       &           &                   &           &        \\
\hline
\multirow{2}{*}{$R_{\rm search} = 100 \ {\bf \rm kpc}$} & & & & & & & & &\\
& & & & & & & & & & \\
\hline
$0.0 < z < 0.5$	&	$29$	&	$0.23$	&	$0.18 \pm 0.05$	 & $0.24$  & $0.10$	&$29$	&	$0.37$	&	$0.20 \pm 0.08$	& $0.45$	& $0.14$ \\
$0.5 < z < 0.8$	&	$142$	&	$0.32$	&	$0.25 \pm 0.03$  & $0.23$  & $0.15$	&$103$	&	$0.62$	&	$0.50 \pm 0.04$	& $0.20$	& $0.56$ \\
$0.8 < z < 1.1$	&	$142$	&	$0.32$	&	$0.26 \pm 0.02$	 & $0.19$  & $0.16$	&$13$	&	$0.54$	&	$0.41 \pm 0.12$	& $0.24$	& $0.38$ \\
$1.1 < z < 1.4$	&	$66$	&	$0.20$	&	$0.12 \pm 0.03$	 & $0.38$  & $0.17$ &       &           &                   &	        &        \\
\hline
\end{tabular}
\label{tab:contamination}
\end{table*}

According to these tests, we obtain that for the vast majority of iterations ($\gtrsim 96$ per cent) the size--mass distributions of paired and isolated massive central galaxies cannot be statistically distinguished. We therefore conclude that the potential contaminants described in this section do not compromise the findings of this work.


\section{Summary and conclusions}\label{sec:conclusion}

Using a compilation of 379 massive (stellar mass M$\gtrsim$10$^{11}\,$M$_\odot$) spheroid-like galaxies from the near-infrared Palomar/DEEP-2 survey,  
we demonstrate that, at least since $z\sim1$, there are not significant differences between the distributions of massive spheroid-like galaxies with (paired) and without (isolated) companions over the size--mass plane. We find that the probability of finding companions around the host galaxy is independent of its size at a given mass, since the companions are not located preferentially around the more compact or extensive massive spheroid-like galaxies. Our finding is independent of the search radius, the redshift and the mass ratio between the spheroid-like massive central galaxy and its companion. 

We explore the size--mass relation for the population of paired and isolated massive spheroid-like galaxies at different redshifts, keeping its slope constant and equal to the obtained by \cite{Shen2003} in the local universe. We analyse the shift between the offsets of the size--mass relation of the paired and isolated populations, finding that are compatible within errors in every case. We also perform two statistical tests, Kolmogorov-Smirnov and t-Student, over both populations to confirm that there do not exist significant differences between them. Given the methodology to identify companions, the uncertainty in the redshift produces a contamination in the fraction of paired galaxies. This uncertainty is independent of the host galaxy position over the size--mass plane, thus we can just correct this effect statistically but not individually. We check that this contaminant factor is not affecting our findings through a Monte Carlo approach.

Our finding suggests that, at least since z$\sim$1, the merger activity in the massive spheroid-like galaxies is rather homogeneous across the whole population and their merger history is not affected by the size of the host galaxy at a given stellar mass. This is very likely suggesting that both compact and extended spheroid-like massive galaxies are growing in size at the same rate. Future studies confronting the observed merger history of massive galaxies with the evolution of the size--mass relation, both its median and intrinsic dispersion, will benefit from the observational constraint presented in this work.

\section*{Acknowledgments}

The authors acknowledge the referee's comments which contributed to improve this work. They also thank Carlos Hernández-Monteagudo for his very useful advices. LADG acknowledges support by the "Caja Rural de Teruel" to develope this research. AJC is a Ram\'on y Cajal Fellow of the Spanish Ministry of Economy and Competitiveness. This work has been supported by the ``Programa Nacional de Astronom\'{\i}a y Astrof\'{\i}sica'' of the Spanish Ministry of Economy and Competitiveness under grants AYA2012-30789, AYA2010-21322-C03-02, AYA2009-10368 and AYA2009-07723-E. This work has made use of the Rainbow Cosmological Surveys Database, which is operated by the Universidad Complutense de Madrid (UCM).


\bibliographystyle{mn2e}
\bibliography{Diaz-Garcia_v1.9}

\end{document}